\documentclass[intlimits,twoside,a4paper]{article}

\usepackage{amsmath,amssymb}
\usepackage{graphicx}
\usepackage{wrapfig}

\usepackage[T2A]{fontenc}
\usepackage[cp1251]{inputenc}

\usepackage[eqsecnum]{cmpj2}

\issue{2016}{19}{2}{23605}
\doinumber{10.5488/CMP.19.23605}

\title[Frustration of freezing in a two dimensional hard-core fluid]%
{Frustration of freezing in a two dimensional hard-core fluid due to particle shape anisotropy\thanks{The authors are pleased to dedicate this article to our good friend and collaborator, Anthony (Tony) Haymet on the occasion of his 60th birthday.  He made a number of valuable contributions to the physics and chemistry of the condensed state.  He has also held a number of important administrative positions.  We wish him well.}
}
\author[A. Huerta, D. Tejeda, D. Henderson, A. Trokhymchuk]{A. Huerta\refaddr{label1},
        D. Tejeda\refaddr{label1}, D. Henderson\refaddr{label2},
        A. Trokhymchuk\refaddr{label3,label4}}
\addresses{
\addr{label1} Facultad de F\'isica, Universidad Veracruzana,
Circuito Gonz\'alo Aguirre Beltr\'an s/n Zona Universitaria, \\
Xalapa, Veracruz, C.P. 91000, M\'exico
\addr{label2} Department of Chemistry and Biochemistry, Brigham Young University,
Provo, UT 84602, USA
\addr{label3} Institute for Condensed Matter Physics of the National  Academy of Sciences of Ukraine, \\ 1~Svientsitskii~St., 79011 Lviv, Ukraine
\addr{label4}
Institute of Applied Mathematics and Fundamental Sciences, Lviv Polytechnic National University, \\ 79013 Lviv, Ukraine
}

\date{Received February 10, 2016, in final form February 25, 2016}
\authorcopyright{A. Huerta, D. Tejeda, D. Henderson, A. Trokhymchuk, 2016}

\begin{document}

\maketitle

\begin{abstract}

The freezing mechanism suggested for a fluid composed of hard disks
[Huerta et al., Phys. Rev. E, 2006, \textbf{74}, 061106]
is used here to probe the fluid-to-solid transition in a hard-dumbbell
fluid composed of overlapping hard disks with a variable length between disk centers.
Analyzing the trends in the shape of second maximum of the radial distribution function of the planar hard-dumbbell fluid
it has been found that the type of transition could be sensitive to the length of hard-dumbbell molecules.
From  the ${NpT}$ Monte Carlo simulations data
we show that if a hard-dumbbell length does not exceed 15\% of the disk diameter,
the fluid-to-solid transition scenario follows the case of a hard-disk fluid, i.e.,
the isotropic hard-dumbbell fluid experiences freezing.
However, for a hard-dumbbell length larger than 15\% of disk diameter, there is evidence that
fluid-to-solid transition may change to continuous transition, i.e., such
an isotropic hard-dumbbell fluid will avoid freezing.
\pacs 64.60.Fr, 68.35.Rh
\end{abstract}

\section{Introduction}

Since Alder and Wainwright \cite{alderwainwright1962} in 1962 showed by molecular dynamics simulations that
two-dimen\-sional (2D) fluid of circular hard-core species~--- hard disks~--- can freeze, a number of papers have been published studying the properties and physics behind such a
phenomenon.
The hard-disk  fluid is one of the simplest interaction models in soft condensed matter science.
This model is far from being trivial as might be thought from a first look and should be
treated in the same way as some other (classical) two-dimensional models such as the Ising model, $XY$ model, and other lattice models. Species of the hard-disk model (e.g., atoms, molecules or particles) do not experience any other interaction except being prohibited from mutual overlap. This apparent simplicity as well as a desire to understand the mechanism that enables the system with no thermal activation to become unstable, undergoing a freezing transition, are the main driving forces of the interest to this model fluid.
The most recent advances in understanding
the physics behind the freezing transition in a hard-disk fluid concern the large scale computer simulation data reported by Zollweg and Chester~\cite{ZollwegPRB1992}, Mak~\cite{MakPRE2006} and Bernard and Krauth~\cite{BernardPRL2011}.
According to these simulation studies, the hard-disk fluid becomes solid in two steps: (i) by means of the first-order type of transition from an isotropic fluid phase to a hexatic phase  and (ii) continuously from a hexatic phase to a solid.

The  monolayers of colloidal particles that very often serve  as a prototype of two-dimensional and quasi-two-dimensional systems (e.g., see review article~\cite{Rice2009} and references therein)
still in various ways (the size polydispersity, shape anisotropy, extra interactions, etc)
differ from the ``ideal'' hard-disk system.
Some of these features, namely,
the extra out-of-core interaction between disks as well as the disk size bidispersity, have been  investigated~\cite{Huerta2004,Huerta2012} and their impact on freezing transition has been revealed.
In particular, it has been shown~\cite{Huerta2012} that in a binary equimolar mixture of hard disks,
there is a limiting large-to-small disk diameter ratio of around 1.2 when freezing transition is still localized, while
the same binary hard-disk mixture having the diameter ratio of around 1.4 does not exhibit the freezing behavior.

In the present study we aim to explore how
the freezing transition in a hard-disk fluid
might be affected by the shape anisotropy imposed on hard-disk species.
In the following section~\ref{sec:2} we start with  the introduction of the shape anisotropy into a hard-disk model and the description of the necessary information regarding the details of computer simulations.
The full set of the results for the present study and discussions are collected in section~\ref{sec:3},
while conclusions are presented in a final section~\ref{sec:4}.

\section{Modelling and details of simulations}
\label{sec:2}

A simple way to introduce shape anisotropy to the planar hard-disk species is to consider the planar hard dumbbell-like objects that consist of two fused hard disks of diameter $\sigma$ and their centers at a distance  $d\sigma$. Such an approach permits to have the hard-core species elongated in one direction with aspect
ratio $(\sigma + d\sigma)/\sigma = 1+d$,  where $d$ plays the role of the anisotropy parameter,
assuming values $0\leqslant d \leqslant 1$.
When $d=0$, the hard-disk fluid is recovered, while the largest anisotropy that can be reached for this hard-core model, $d=1$, corresponds to the case of tangent hard disks and will be referred to as a hard-dimer fluid.
Besides the anisotropy parameter, the system is characterized by packing fraction $\eta$ defined as $\eta=\rho A_{\rm d}$, where $\rho = N/A$ and  $A$ is the area of the system,
 $A_{\rm d}$ is the area of a hard-dumbbell molecule that depends on the anisotropy parameter $d$, while $N$ is the number of hard-dumbbell species.
\begin{figure}[!b]
\includegraphics[width=0.465\textwidth]{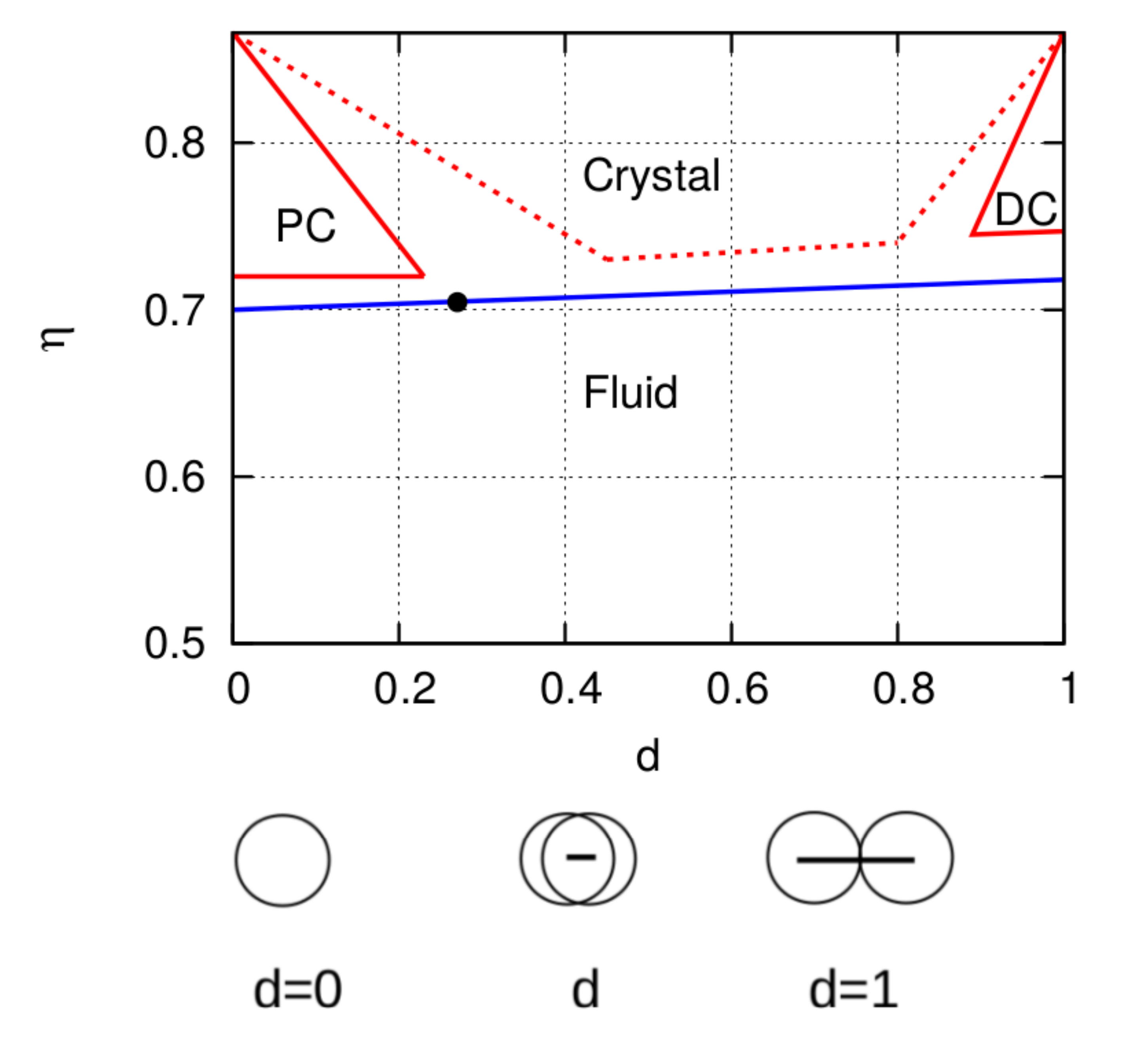}
\includegraphics[width=0.45\textwidth]{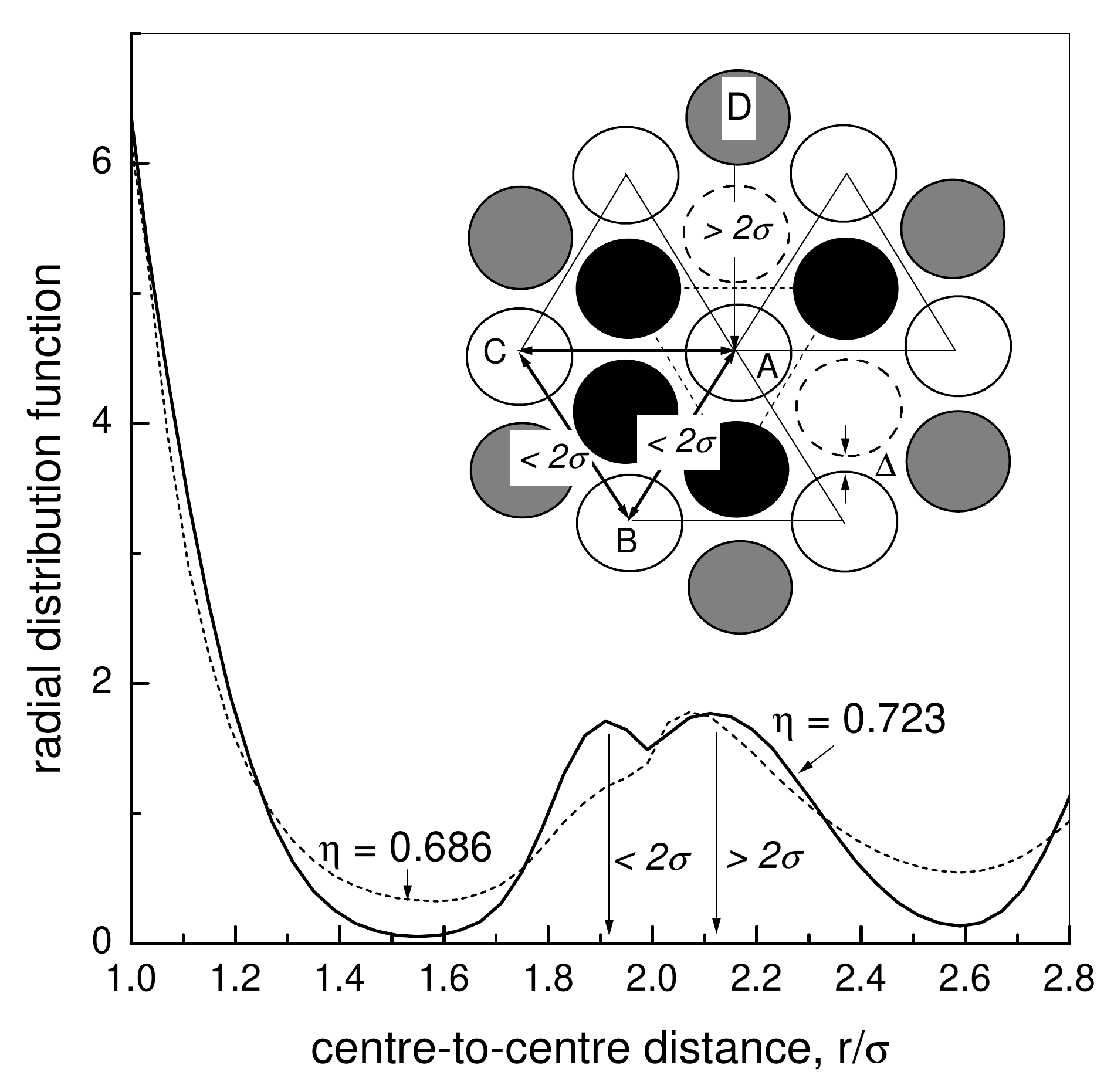}
\caption{(Color online) Left: Phase diagram of the hard-dumbbell system suggested by Wojciechowski~\cite{WojciechowskiPLA1987},  and schematic of a single hard-disk molecule of diameter $\sigma$, and hard-dumbbell molecules that consists of two fused hard disks of diameter $\sigma$ and centers at a distance  $d\sigma$.
Right: Monte Carlo simulations data for the radial distribution function of a hard-disk system at two packing fractions $\eta$: before freezing
(dashed line) and before melting (solid line). Inset: Sketch of the disk regular hexagonal configurations where four disks filled with black and two empty disks
drawn by the dashed line are the first coordination shell neighbors of the central disk $A$, while those six filled with gray and those six empty disks drawn by the solid line
belong to the second shell neighbors.}
  \label{freez}
\end{figure}

Figure~\ref{freez} (the left-hand-side frame) shows the hard-dumbbell modelling as well as the phase diagram  in anisotropy-density coordinates that
was suggested by Wojciechowski~\cite{WojciechowskiPLA1987} based on heuristic reasoning using free-volume arguments. According to this diagram,
any hard-dumbbell system will always undergo a
first-order transition from an isotropic fluid phase to a solid phase. The latter could be of various types such as an ordinary crystal, a plastic crystal (PC) or a disordered crystal (DC) depending on the magnitude of the anisotropy parameter $d$.
As it has been already mentioned in the Introduction, a scenario that involves first-order transition indeed takes place in the limit $d=0$ case, i.e., for a hard-disk fluid~\cite{ZollwegPRB1992,MakPRE2006,BernardPRL2011}.
The transition with two coexisting phases~--- isotropic fluid and disordered crystal phase~--- has been also confirmed  for another limiting case $d=1$, i.e., for a hard-dimer system~\cite{WojciechPRL1991}, as well as for a hard-dumbbell fluid with  $d < 1$, namely, in the case of $d=0.924$~\cite{WojciechPRB1992}.

In the present study, we consider the features of fluid-to-solid transition in a hard-dumbbell fluid that is characterized by a small deviation of molecule shape anisotropy from a hard-disk fluid, by assuming the values of the anisotropy parameter within the range $0\leqslant d < 0.28$.
The latter was chosen by means of the anisotropy-density phase diagram in figure~\ref{freez} that associates this range of a hard-dumbbell fluid anisotropy with the coexisting plastic crystal solid.

We have used Monte Carlo simulations technique  to study the properties related to freezing transition in a  hard-dumbbell fluid.
All simulations were performed in a rectangular box with an aspect ratio of $\sqrt{3}/2$ with the usual  periodic boundary conditions.
Two different thermodynamic ensembles have been employed.
Namely, the constant-density ${NVT}$ ensemble was used to evaluate the radial distribution function  and global bond-orientational order parameter of the hard-dumbbell fluid,
while the constant-pressure  ${NpT}$ ensemble was used
to discriminate between a first-order and continuous transition in the same system.
The symbols $N, V, T$ and $p$ specify the
number of particles, volume (area in two dimensions), temperature and pressure
used in the simulations, respectively.

In the course of our simulations, the single molecular move (in what follows is referred to as the iterative step) consists of a random positional displacement of molecular mass center accompanied by a rotation of molecular axes. For denser states, we also attempted a $\pi/3$ orientational displacement of molecular axes that may be commensurable with a crystalline solid phase.
The ${NpT}$ simulations additionally consist in
the change of the area of the simulation box.
The acceptance ratio was maintained
between 20\% and 30\%  by adjusting the maximal size of positional displacement and maximal amplitude for the rotation in the case of ${NVT}$ ensemble, and between 20\% and 50\%  by adjusting the maximal size of positional displacement, maximal amplitude for the rotation and maximal variation of the box size in the case of ${NpT}$ ensemble.

A standard Metropolis algorithm was used to obtain the ensemble averages.
For ${NVT}$ simulations, each equilibration run was relaxed for at least  $10^4$ iterative steps.
The resulting data for radial distribution function and  global bond-orientational order parameter were obtained by averaging at least over 600 configurations, each being relaxed by $10^3$ iterative steps.
Whereas for the ${NpT}$ simulations, each equilibration run was relaxed for at least $10^6$ iterative steps,
the distribution of densities was accumulated after at least $2\times 10^7$
iterative steps to obtain a good sampling.

Various sizes of system were investigated, ranging from 100 to 1000 molecules.
Similar to the case of a hard-disk fluid~\cite{TruskettPRE1998,Huerta2006}, no systematic system size effects were observed in the ${NVT}$ data for the second maximum of
the radial distribution function, and subsequent calculations were made using $N =400$ hard-dumbbell molecules. These simulations have been used to collect data for the global bond-orientational order parameter. However, in this case
it is well documented~\cite{Weber} that $N =400$ is too small to identify
the boundary of freezing and melting densities. Therefore, the data for global bond-orientational order parameter could be used for qualitative purposes only.

The Monte Carlo simulations in the ${NpT}$ ensemble are
efficient at identifying the type of fluid-to-solid transition.
According to the studies by Lee and Strandburg~\cite{Strandburg1992} on the application of the isobaric Monte Carlo simulations, the system size $100\leqslant N \leqslant 400$ is sufficiently large to identify the fluid-to-solid transition in a hard-disk system as first order transition.
Therefore, the ${NpT}$ calculations of the equation of state were made using $N =100$ hard-dumbbell molecules. The consequences  of such a system size for pressure magnitude of the hard-disk fluid in the vicinity of transition region are already discussed in the literature~\cite{Kolafa}.

\clearpage

\section{Results and discussions}
\label{sec:3}

Following the observation by Truskett et al.~\cite{TruskettPRE1998}, the fluid composed of planar hard disks (the case of the anisotropy parameter $d=0$) exhibits a structural precursor to freezing transition that manifests itself as a shoulder in the second
maximum of the disk-disk radial distribution function.
In fact, an increase of the density in a planar array of $N$ hard disks of diameter $\sigma$, that are uniformly spread on area $A$, is accompanied by formation of the local quasi-regular hexagonal arrangement.
By quasi-regular hexagons we mean the sixfold configurations (see the right-hand side frame of figure~\ref{freez} adopted for details from reference~\cite{Huerta2006}) that, at certain disk density or  packing fraction $\eta=\pi N\sigma^2/(4A)$, is characterized by a constant gap width
%
$$\frac{\Delta}{\sigma} = \left(\frac{\eta_{\rm cp}}{\eta}\right)^{1/2}-1$$
%
between each pair of the neighbor disks~\cite{Huerta2006}.
Here, $\eta_{\rm cp}=\pi/ (2\sqrt{3})\approx 0.907$ is the disk close packing (CP) fraction. Such a sixfold formation implies that center-to-center distances $r$  between disks and, consequently, the gap width $\Delta$ shortens under the density increasing.

By analyzing the schematic drawing in the inset of figure~\ref{freez},
Huerta et al.~\cite{Huerta2006}  pointed out that there exists a packing fraction $\eta_{\rm cage} = \pi\sqrt{3}/8 =0.680$ and, consequently, a gap width $\Delta/\sigma =2/\sqrt{3}-1\approx 0.155$, which in transparent way  demonstrates the existence
of two kinds of the
second shell neighbors discerned by means of the distance criterion. Namely, the second shell neighbors   on the distances (e.g., AD,~\ldots)
that are always larger than $2\sigma$,  and those  on the distances (e.g., AB, AC,~\ldots) that could be shorter than $2\sigma$.
In fact, the existence of these two kinds (larger and shorter) of the second shell neighbors is the origin, initially of a shoulder and then, with an increase of the density, a split of the second peak of the disk-disk radial distribution function $g(r)$ for distances around $r/\sigma=2$.

On the other hand, it implies that the center-to-center distances $r$ between the  first-shell alternating neighbors of any disk in a fluid system become, on average, shorter than $2\sigma$, {i.e.,} the set of only three first-shell alternating neighbors is capable of forming a cage, preventing the central disk to wander. Thus, the hard-disk fluid will exhibit a tendency to freeze.
The corresponding hard-disk fluid caging density,
$\eta_{\rm cage} =0.680$, is
lower than the freezing transition density, $\eta_{\rm f} = 0.700$, that follows from large scale computer simulation data~\cite{BernardPRL2011}.
Actually it means that in practice disks are not spread uniformly and
some of the center-to-center distances (AB, AC,~\ldots) become shorter than $2\sigma$ already for the disk packing fractions $\eta \geqslant 0.6$, initiating deformation of the smooth second maximum of the disk-disk radial distribution function~$g(r)$.

Therefore, a starting point to discuss the effect of shape anisotropy on the freezing transition in a fluid composed of planar hard dumbbells
will be to analyze the trends in the second
maximum of the center-center radial distribution functions shown in figure~\ref{freez} for  packing fractions $\eta=0.686$ and $0.723$ upon increasing the value of anisotropy parameter $d$.

\begin{figure}[!t]
 \centering
\includegraphics[width=0.495\textwidth]{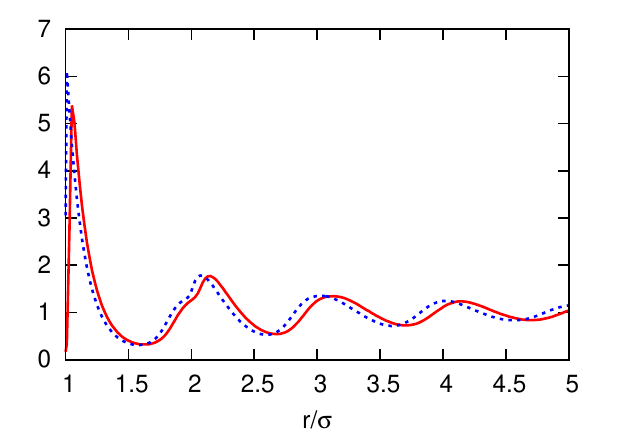}
\includegraphics[width=0.495\textwidth]{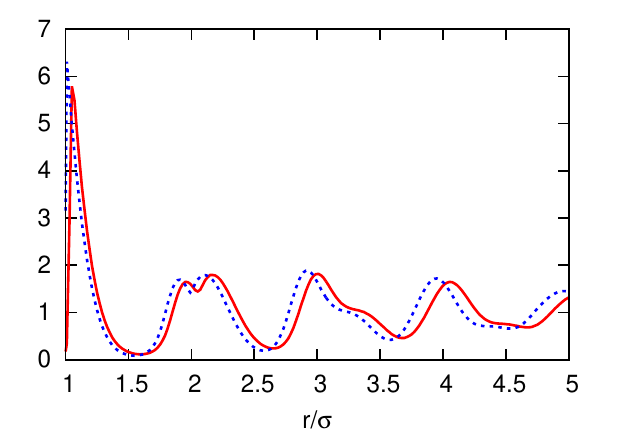}
 \caption{(Color online) The ${NVT}$ Monte Carlo simulations data for the center-center radial distribution function $g(r)$ of the hard-dumbbell fluid with anisotropy parameter $d=0.05$ at two values of packing fraction:  $\eta=0.686$, associated with an isotropic fluid phase (the left-hand-side frame) and $\eta=0.723$, associated with an ordered solid phase (the right-hand-side frame). Both frames show a comparison against the corresponding radial distribution function of the hard-disk, $d=0$, fluid (the dotted line).
 }
 \label{rdfa}
\end{figure}

\begin{figure}[!b]
 \centering
\includegraphics[width=0.495\textwidth]{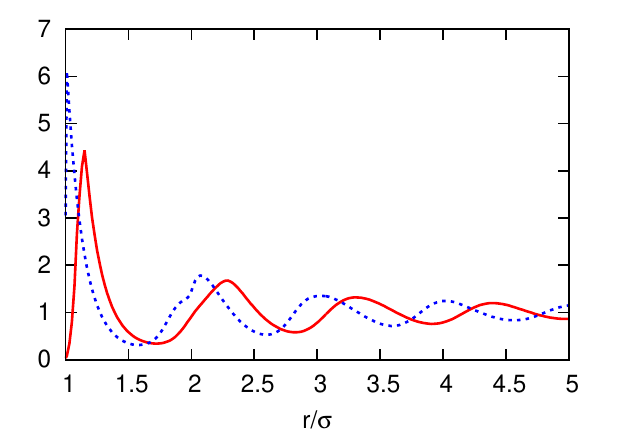}
\includegraphics[width=0.495\textwidth]{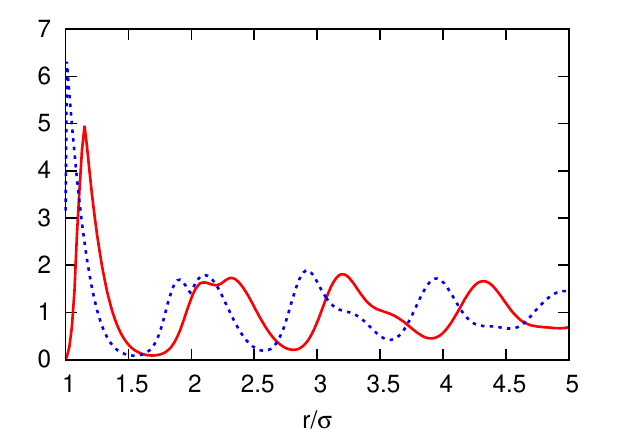}
\caption{(Color online) The same as in figure~\ref{rdfa} but for the anisotropy parameter $d=0.15$.}
\label{rdfb}
\end{figure}

\begin{figure}[!t]
 \centering

\includegraphics[width=0.495\textwidth]{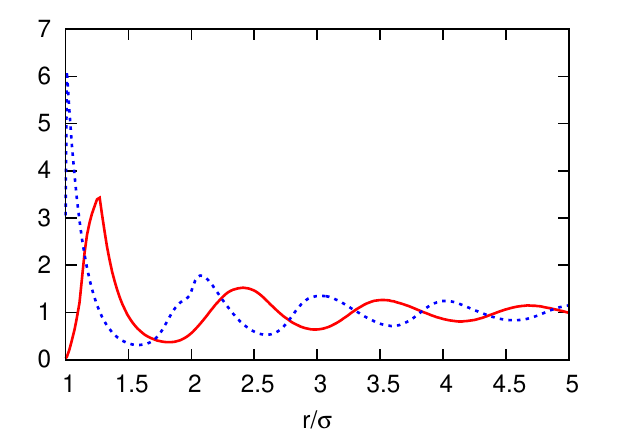}
\includegraphics[width=0.495\textwidth]{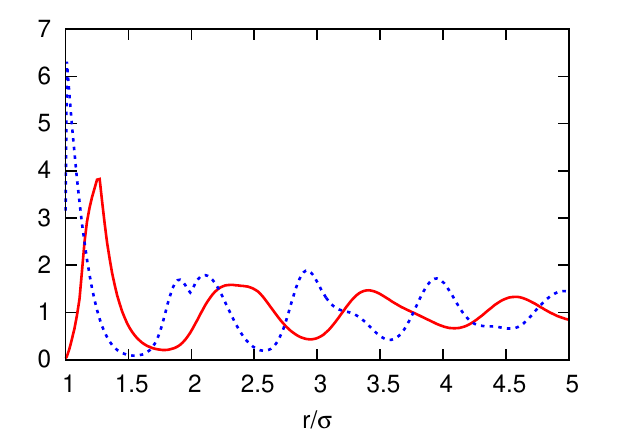}
\caption{(Color online)  The same as in figure~\ref{rdfa} but for the anisotropy parameter $d=0.27$.}
 \label{rdfc}
\end{figure}

Figures~\ref{rdfa}--\ref{rdfc} show a set of ${NVT}$ Monte Carlo simulation data for the center-center radial distribution functions $g(r)$
of the hard dumbbell fluids characterized by three values of the shape anisotropy parameter
$d = 0.05$, 0.15 and  0.27.
that are explicitly compared against the case of hard disk fluid with $d = 0$.
In each case, the radial distribution functions are shown
at two values of packing fraction, $\eta = 0.686$ (left-hand side panel) and $\eta = 0.723$ (right-hand side panel).
These particular values of packing fractions correspond to the close proximity
of the freezing and melting transitions, respectively, in the case of hard-disk fluid ($d=0$).
To facilitate the comparison and discussion, in all cases that are presented in figures~\ref{rdfa}--\ref{rdfc},
the data for hard-dumbbell fluid are explicitly compared against the corresponding data for hard-disk fluid.


As expected, and in agreement with a prediction of the density-anisotropy phase diagram
in figure~\ref{freez},
for  small  values of the anisotropy parameter, $d=0.05$, the freezing scenario seems to be quite similar to the case of a hard-disk fluid.
Namely, for the case of $d=0.05$ we are observing practically the same shape of the second
maxima for both densities with a tiny  shift in the positions to slightly larger center-to-center distances $r$.

As the shape anisotropy increases to $d=0.15$, the shift in the positions of maxima and minima of the $g(r)$ becomes more pronounced. Although for packing fraction $\eta=0.686$, the shoulder in the second maxima is still noticed but it is less evident. However, since the split of the second maxima is definitely present, one could expect that the freezing mechanism that was discussed for a hard-disk fluid is still at work.

However, the case of the shape anisotropy parameter $d=0.27$ seems to be crucial from the point of view of the freezing transition. The resulting radial distribution function in such a hard-dumbbell fluid, shown in
figure~\ref{rdfc}, is totally different from the one of the hard-disk fluid for both packing fractions considered. Most important is that there are no indications  either of a shoulder at lower packing fraction or of a split of the second maximum at higher packing fraction.
When discussing figure~\ref{freez}, we have already pointed out that besides serving as a precursor towards approaching the freezing transition, the presence of these features is signaling about the
formation of the local quasi-regular hexagonal arrangement.

\begin{figure}[!b]
\centering
\includegraphics[width=0.55\textwidth]{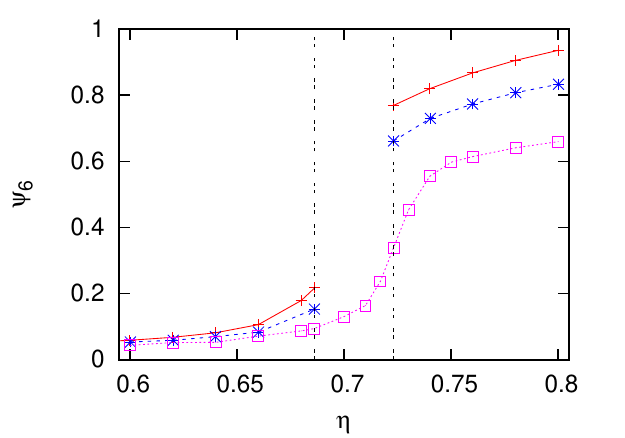}
\caption{(Color online) The ${NVT}$ Monte Carlo data for global bond-orientational order parameter $\psi_6$ of a hard-dumbbell fluid
with anisotropy parameter values $d = 0.27$ and 0.15 (from the bottom to the top) in comparison against a hard-disk ($d=0$) fluid system (the solid line at the top).
The vertical dashed lines mark the fluid-to-solid transition region that corresponds to the system with $N=400$.
} \label{psi6}
\end{figure}

A quantitative measure of hexagonal order  in the system is provided by the so-called
global bond-orientational order parameter $\Psi_6$, that was
evaluated during the ${NVT}$ Monte Carlo simulation runs using
definition
 \begin{equation}
 \Psi_6 = \Bigg|\frac{1}{N_\text{nn}}\sum_{j=1}^{N}\sum_{k=1}^{N_\text{nn}}
 \re^{\displaystyle{\ri6\theta_{jk}}}\Bigg|,
 \nonumber
 \end{equation}
where $j$ runs over all disks in the system, $k$ runs over all
geometric nearest neighbors (nn) of disk $j$, each obtained
through the Voronoi analysis and $N_\text{nn}$ is the total number of
such nearest neighbors in the system. The angle $\theta_{jk}$ is defined
between some fixed reference axis in the system and
the vectors bonds connecting nearest neighbors $j$ and $k$.
As one can see from figure~\ref{psi6}, indeed,  the global bond-orientational order parameter $\psi_6$ of the hard-dumbbell fluid with anisotropy parameter $d=0.27$ being very small at low packing fraction, continuously grows under density increasing towards a denser phase.

\begin{figure}[!t]
\centering
\includegraphics[width=0.495\textwidth]{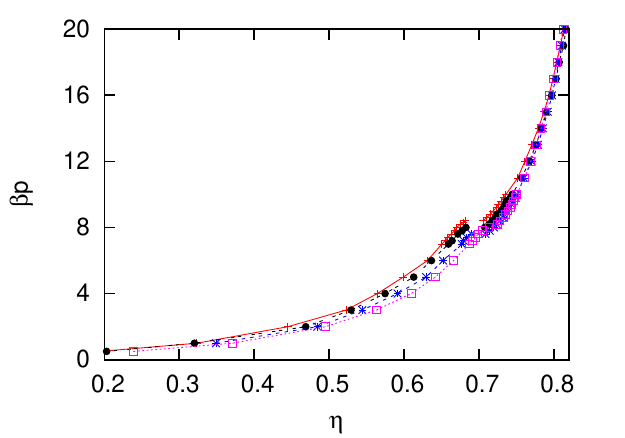}
\includegraphics[width=0.495\textwidth]{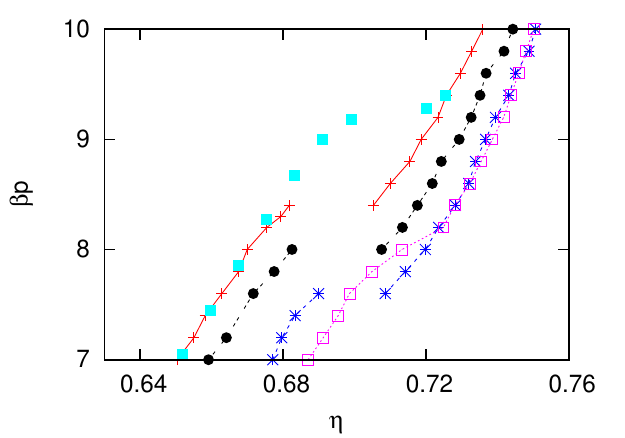}
\caption{(Color online)  The ${NpT}$ Monte Carlo data for pressure $\beta p$  of a hard-dumbbell fluid with anisotropy parameter values $d = 0.27$, 0.15 and 0.05 (from the bottom to the top for $\eta < 0.7$) in comparison against a hard-disk ($d=0$) fluid (at the top). The symbols correspond to Monte Carlo simulation data while lines are shown to guide the eyes.
The left-hand-side frame presents the full density range, while the right-hand-side panel shows the region nearby the  fluid-to-solid transition only.
In the right-hand-side panel: the vertical dashed lines mark the fluid-to-solid transition region according to Bernard and Krauth~\cite{BernardPRL2011};
the filled squares at the top represent computer simulation data for a hard-disk fluid that are corrected for finite-size effects~\cite{ZollwegPRB1992,alder1968,Kolafa}.
} \label{eos}
\end{figure}

Analyzing the trends in the second maximum of the radial distribution function in  figures~\ref{rdfa}--\ref{rdfc}, as well as the trends in profile of the global bond-orientational order parameter in figure~\ref{psi6}, one can expect that the hard-dumbbell fluid with anisotropy parameter  $d > 0.15$ could behave differently than the hard-disk fluid is doing  while approaching the fluid-to-solid transition region.
To obtain a better insight into the features of fluid-to-solid transition in the system composed of hard-dumbbell molecules,
in figure~\ref{eos} we show the ${NpT}$ Monte Carlo data for pressure $\beta p$, where $\beta=1/k_{\rm B}T$ and $k_{\rm B}$ is the Boltzmann constant, in a wide range of packing fraction $\eta$ (the left-hand-side frame of figure~\ref{eos}).
 The hard-dumbbell systems presented in figure~\ref{eos} are the same
as have been discussed in figures~\ref{rdfa}--\ref{psi6} to learn about the trends in the features of the second maximum of the radial distribution function.
As expected, those features that are sensitive to the shape anisotropy in the density region that precedes fluid-to-solid transition; for better visualization, this region is  shown separately on the right-hand-side frame of  figure~\ref{eos}. On the other hand, from the results in the left-hand-side frame of figure~\ref{eos} it follows that in the limiting case of low ($\eta < 0.3$) as well as at high ($\eta\approx 0.8$) packing fractions, all pressure isotherms tend to coincide, showing no dependence on the anisotropy parameter.

Although quantitatively the results for pressure $p$ and for the freezing and melting transition densities could be affected by the system size effects, which is rather evident in the case of a hard-disk fluid, the qualitative picture must be correct. Namely, like in the previous ${NpT}$ Monte Carlo simulations~\cite{Strandburg1992} as well as in quite recent   large-scale ${NVT}$ Monte Carlo simulations~\cite{BernardPRL2011,EngelPRE2013,KindtJCP2015} of the freezing transition in a hard-disk system, our ${NpT}$ Monte Carlo data for a hard-disk ($d = 0$) fluid show a discontinuous first-order transition from an isotropic fluid phase to a denser and highly ordered phase. The same scenario persists for hard-dumbbell fluids with anisotropy parameter values $d = 0.05$ and 0.15.
However, in the case of a hard-dumbbell fluid with anisotropy parameter $d = 0.27$, a continuous transition is obtained.
Regarding the issue of the system size effects in the case of a hard-dumbbell fluid, we note that contrary to a hard-disk system, all the existing computer simulation data for a hard-dumbbell system have been obtained with $N =112$ molecules~\cite{WojciechowskiPLA1987,WojciechPRL1991,WojciechPRB1992}.

\begin{figure}[!b]
\centering
\includegraphics[width=0.33\textwidth,height=0.35\textwidth]{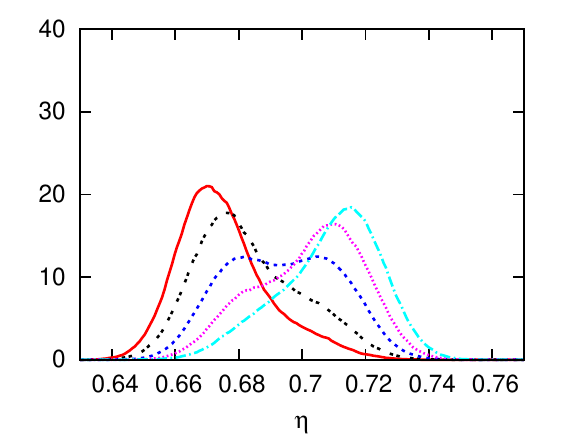}
\includegraphics[width=0.33\textwidth,height=0.35\textwidth]{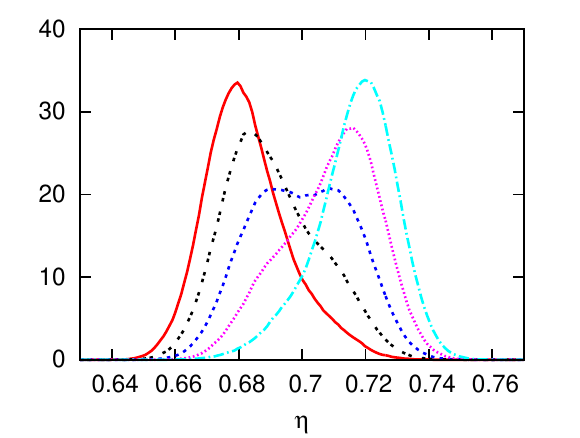}
\includegraphics[width=0.33\textwidth,height=0.35\textwidth]{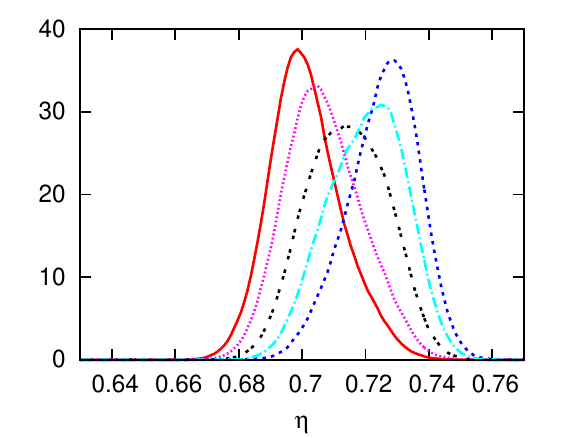}
\caption{(Color online) Density distributions obtained in the ${NpT}$ Monte Carlo simulations of the
system of $N=100$ hard-dumbbell molecules with three different values of the anisotropy parameter. Namely, $d=0$, i.e., for a hard-disk fluid and for two hard-dumbbell fluids with $d=0.15$ and 0.27 from the left to the right.
Each curve for a particular system corresponds to a constant pressure $p$ shown by symbols in  figure~\ref{eos}.
}
\label{dist}
\end{figure}

In fact, the statement regarding the frustration of freezing in a hard-dumbbell fluid with the aniso\-tropy parameter $d=0.27$ is consistent with the information that follows from the data presented in figure~\ref{dist} where density distributions obtained in the ${NpT}$ Monte Carlo simulation are shown.
Each curve for a particular system in figure~\ref{dist} corresponds to a constant pressure $p$, and the position of its maximum determines the density (packing fraction $\eta$) of the stable phase of this system in this thermodynamic state. The curves with two maxima in the case of $d=0$ and $d=0.15$ each tells us that at pressure $p$, that corresponds to this curve, there is coexistence of two phases and, consequently, the fluid-to-solid transition is  discontinuous or of first order.  By contrast, in the third case of hard-dumbbell fluid with $d=0.27$, all curves show only one maximum, and under pressure increasing, the position of this maximum is moving towards higher densities (packing fractions $\eta$). Thus, the fluid-to-solid transition in this case is continuous.

\section{Conclusion}
\label{sec:4}

Contrary to the hard-disk system,
where fluid-to-solid transition has been the subject of a long-stan\-ding debate~\cite{Strandburg1988}, the hard-dumbbell system has not been explored in full yet.
This fact looks rather strange in view that almost three decades ago the theoretically predicted phase diagram of the hard-dumbbell system in full range of the anisotropy parameter, $0\leqslant d \leqslant 1$, was reported based on heuristic reasoning and using free-volume arguments~\cite{WojciechowskiPLA1987}. The main conclusion of that study was that in the full range of the anisotropy parameter values, fluid-to-solid transition is of the first order.
Up to date, this result was confirmed by computer simulations for two particular values of the anisotropy parameter, i.e., $d=1$~\cite{WojciechPRL1991} and 0.924~\cite{WojciechPRB1992}. Obviously, theoretical predictions are valid for the case of $d = 0$, i.e., for a hard-disk fluid~\cite{MakPRE2006,Weber,BernardPRL2011,EngelPRE2013,KindtJCP2015}.

In this study, we turned our attention to the case of anisotropy parameter values range $0\leqslant d \leqslant 0.28$ that was indicated in that theoretical analysis~\cite{WojciechowskiPLA1987} as a unique one where the isotropic fluid phase freezes by coexisting with the plastic crystal solid.
First of all, using the ${NVT}$ Monte Carlo simulations data we have shown that such a feature of the radial distribution function as the shoulder in its second maximum, that was suggested by Truskett et al.~\cite{TruskettPRE1998} as the structural precursor to freezing transition in the case of a hard-disk fluid,  is still preserved for a hard-dumbbell fluid with anisotropy parameter values up to $d\leqslant 0.15$. However, in the case of a hard-dumbbell system with anisotropy parameter $d = 0.27$, the radial distribution functions do not show any sign of the shoulder in the range of distances that correspond to the second maximum.
Secondly, by performing the ${NpT}$ Monte Carlo simulations we obtained that indeed first-order freezing transition from an isotropic fluid to denser highly ordered phase persists in the case
of a hard-dumbbell fluid with an anisotropy parameter $d\leqslant 0.15$ while two-phase coexistence is absent in the case of $d = 0.27$.

Although in the present study we did not consider this question, the coexisting denser phase in the case of a hard-dumbbell fluid, with anisotropy parameter $d\leqslant 0.15$,  could be associated with a plastic crystal, as it was predicted~\cite{WojciechowskiPLA1987}. We remind the reader that the plastic crystal phase  represents one where densely packed molecules  are still free to rotate.
In fact, close examination of the hard-disk fluid configurations in the vicinity of freezing transition revealed the formation of a local quasi-regular hexagonal arrangement of hard disks~\cite{Huerta2006}.
Under increasing density, the average center-to-center distance between neighboring disks continuously shortens and, at the point of freezing. This results in a disk spacing of approximately 15\% of disk diameter.
Evidently, any disturbance that will tend to violate such a spacing, e.g., due to the attractive or repulsive interaction between disks, or the degree of disk size polydispersity, {etc.}, could potentially result in the frustration of the freezing phenomenon.
This is what we have observed in the present study: if the disk elongation starts to
exceed 15\% of the disk diameter (the anisotropy parameter $d=0.27$), the freezing transition cannot survive and under an increase of the density, the disordered isotropic hard-dumbbell fluid undergoes a continuous transition to a denser and more ordered phase.
Such a mechanism has already proved to work in the case of hard disks with square-well attraction~\cite{Huerta2004} and for an equimolar binary hard-disk mixture~\cite{Huerta2012}.

\section*{Acknowledgements}

This work is supported by the CONACYT under the project 152431,
Red Tem{\'a}tica de la Materia Condensada Blanda and Promep of M{\'e}xico.

\ukrainianpart

\title{Уникнення замерзання двовимірного плину жорстких молекул через анізотропію їх форми}
\author{А. Уерта\refaddr{label1}, Д. Техеда\refaddr{label1}, Д. Гедерсон\refaddr{label2},  А. Трохимчук\refaddr{label3,label4}}
\addresses{
\addr{label1} Унiверситет Веракрузана, факультет фiзики та iнженерiї, кафедра фiзики, \\
Халапа, Веракруз, СР 91000, Мексика
\addr{label2} Факультет хімії та біохімії, Університет Брігама Янга, Прово, Юта 84602, США
\addr{label3} Iнститут фiзики конденсованих систем НАН України,
вул. І.~Свєнцiцького,~1, 79011 Львiв, Україна
\addr{label4} Інститут прикладної математики та фундаментальних наук, Національний університет ``Львівська Політехніка'',  79013 Львів, Україна
}

\makeukrtitle

\begin{abstract}
\tolerance=3000%

Механізм замерзання, який був запропонований для плину жорстких дисків [Huerta A. et al., Phys. Rev. E, 2006, \textbf{74}, 061106], використовується тут для дослідження фазового переходу плин-тверде тіло  у системі, що складається з жорстких гантелеподібних молекул, утворених двома жорсткими дисками зі змінною відстанню між їх центрами. З аналізу тенденцій зміни форми другого максимуму радіальної функції розподілу зроблено висновок, що тип фазового переходу плин-тверде тіло може бути чутливий до видовження гантелеподібних молекул.  На основі даних комп'ютерного експерименту Монте Карло при фіксованому тиску було знайдено, що коли видовження молекул не перевищує 15\% діаметра жорсткого диска, то фазовий перехід плин-тверде тіло відбувається за тим же сценарієм що і у системі жорстких дисків, тобто плин гантелеподібних молекул замерзає. У випадку, коли видовження молекул перевищує 15\% діаметра жорсткого диска, то є  підстави стверджувати, що фазовий перехід плин-тверде тіло проходить неперервно, тобто плин гантелеподібних молекул у цьому випадку не замерзає.

\keywords плин жорстких дисків, гантелеподібні молекули, радіальна функція розподілу, \\ фазовий перехід замерзання

\end{abstract}

\end{document}